\documentclass{article}
\usepackage{spconf,amsmath,graphicx}
\RequirePackage{booktabs}
\usepackage{multirow}
\usepackage{amssymb}
\usepackage{pifont}
\usepackage{cite}

\DeclareMathOperator*{\argmin}{argmin}

\title{Efficient black-box speaker verification model adaptation with reprogramming and backend learning}
\name{Jingyu Li, Tan Lee}
\address{Department of Electronic Engineering, The Chinese University of Hong Kong, Hong Kong}
\begin{document}
\ninept
\maketitle
\begin{abstract}
The development of deep neural networks (DNN) has significantly enhanced the performance of speaker verification (SV) systems in recent years. However, a critical issue that persists when applying DNN-based SV systems in practical applications is domain mismatch. To mitigate the performance degradation caused by the mismatch, domain adaptation becomes necessary. This paper introduces an approach to adapt DNN-based SV models by manipulating the learnable model inputs, inspired by the concept of adversarial reprogramming. The pre-trained SV model remains fixed and functions solely in the forward process, resembling a black-box model. A lightweight network is utilized to estimate the gradients for the learnable parameters at the input, which bypasses the gradient backpropagation through the black-box model. The reprogrammed output is processed by a two-layer backend learning module as the final adapted speaker embedding. The number of parameters involved in the gradient calculation is small in our design. With few additional parameters, the proposed method achieves both memory and parameter efficiency. The experiments are conducted in language mismatch scenarios. Using much less computation cost, the proposed method obtains close or superior performance to the fully finetuned models in our experiments, which demonstrates its effectiveness.



\end{abstract}
\section{Introduction}
\label{sec:intro}

A typical speaker verification (SV) system consists of two major processes, extracting speaker embeddings and similarity computation on the embeddings\cite{liu2015deep,xie2019utterance,chung20b_interspeech,10094954}. A high similarity between two utterances' speaker embeddings indicates that they belong to the same speaker, conversely, different speakers. Deep neural network (DNN) - based SV systems have demonstrated great performance and have been widely used recently\cite{snyder2017deep,snyder2018x,desplanques20_interspeech}. Domain mismatch commonly appears in practical applications, e.g., language mismatch and various background sounds. The existence of mismatches depresses the verification accuracy of a pre-trained SV model significantly. To alleviate this problem, different domain adaptation methods are investigated to retain the performance and robustness of the models\cite{lin2020framework,wang2018unsupervised,chen2020adversarial,alam2018speaker,li22m_interspeech}. 

\begin{figure}[t]
  \centering
  \includegraphics[width=\linewidth]{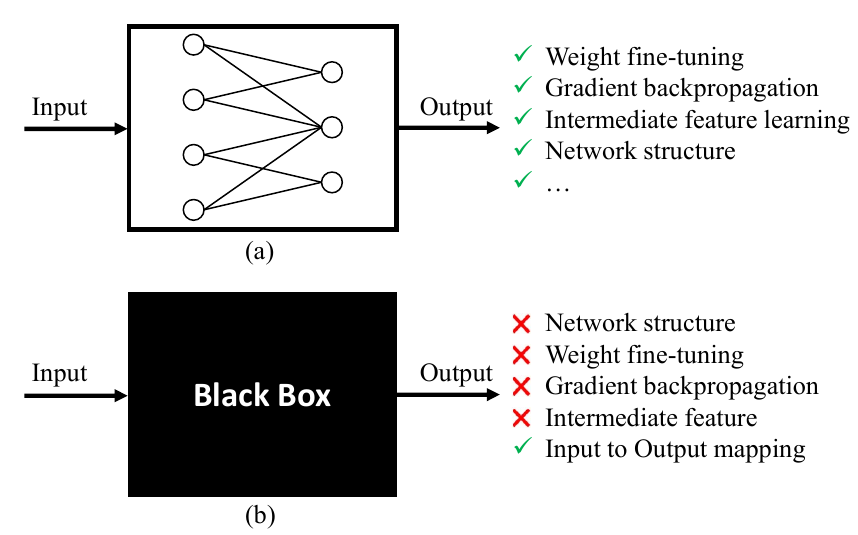}
  \caption{The illustration of (a) a white-box model and (b) a black-box model.}
  \label{fig:blackbox}
  \vspace{-2.5mm}
\end{figure}

One straightforward strand is finetuning the pre-trained speaker embedding extraction models with target domain data\cite{rohdin2019speaker,lin2020framework}, which helps the model learn new knowledge from the target domain. Adversarial training was introduced in SV adaptation\cite{chen2020adversarial,wang2018unsupervised,rohdin2019speaker} to decrease the domain-variant factors of the model so that the model could perform solidly under different domains. The model finetune is computationally costly on a large-scale model. Moreover, these methods require accessing the parameters and architecture of the pre-trained models, however, which may not be achieved in some scenarios.
For instance, a commercial DNN model could be packaged as a black box without exposing the details of the model. The black-box models are common in practical applications, such as APIs and some software. The model only gives the input-output reaction but nothing else, as shown in Fig.~\ref{fig:blackbox}. In this situation, model finetuning is not available. 

DNN-model backend output has been utilized in transfer learning for downstream tasks in many studies\cite{ren2015faster,hsu2021hubert}. Another trend of adaptation learning is based on the backend speaker embeddings from the pre-trained SV models\cite{li22m_interspeech,alam2018speaker,li2022coral++}, e.g., CORAL and EDITnet. Instead of high-dimensional data like an input waveform or spectrum, the speaker embedding is a simple vector whose length is usually less than $1000$. The embedding-based adaptation methods are lightweight without modifying the pre-trained model and are available for black-box models. The drawback of the backend learning approach is obvious. More of the input information has been lost in the speaker embeddings, thus it may not be sufficient for achieving a good result.

In this paper, we propose a novel domain adaption method for black-box SV models by augmenting backend learning with reprogramming\cite{elsayed2018adversarial}. Reprogramming was utilized in transfer learning in various fields\cite{tsai2020transfer,yang2021voice2series,hung2023low,yang2023english,yen23_interspeech}. Reprogramming modifies a model's output by changing the model's input with learnable parameters, which keeps the pre-trained model unchanged, as shown in Fig.~\ref{fig:network}(c). The update of the input-level learnable parameters requires gradient backpropagation through the model, thus vanilla reprogramming can not be applied to black-box models. To address this problem, we propose to utilize a neural network for gradient estimation for the model input. The gradient estimation network is designed to be small, so the gradient calculation during training is lightweight. After training, the gradient estimator is abandoned and only a few parameters are added to this SV model.

By reprogramming the input, the output speaker embedding is modified to include more information, which is beneficial for the following backend learning. We investigate the proposed method for language adaptation on pre-trained SV models. Our experiments are conducted on different scales of data with different model structures. The results show that the proposed method decreases the performance degradation under language mismatch with low computation cost and few parameter additions. It outperforms the full finetuned model in some networks.


\section{Reprogramming}
\label{sec:reprog}
Adversarial reprogramming was first proposed in \cite{elsayed2018adversarial}. The reprogramming consists of two major steps\cite{yang2021voice2series}: $(1)$ manipulating the model input and $(2)$ mapping the output labels from the source task to the target task. The common input manipulation method is concatenating\cite{yang2021voice2series,tsai2020transfer} or adding\cite{hung2023low,yang2023english,yen23_interspeech} learnable parameters with the original model, which can be formulated as\cite{yang2021voice2series}:
\begin{equation}
    \Tilde{x} = zero\_padding(x) + W \odot M
  \label{eq:reprog}
\end{equation}
where $\mathbf{W}$ is the learnable parameters and $M$ is a mask function. If the mask equals $1$, this equation represents an addition operation. If the mask equals $0$ on the region of original input $\mathbf{X}$ and $1$ otherwise, it works as concatenating new parameters with $\mathbf{X}$. 

\cite{elsayed2018adversarial} has shown that a pre-trained DNN model can perform other tasks different from its original one by reprogramming, which is suitable for transfer learning. \cite{yang2021voice2series,hung2023low,yang2023english,yen23_interspeech} show the effectiveness of reprogramming in transfer learning in audio and speech fields, e.g., reprogramming acoustic models for time series classification, music genre classification, and cross-lingual speech recognition.

\begin{figure}[t]
  \centering
  \scalebox{1.02}{
  \includegraphics[width=\linewidth]{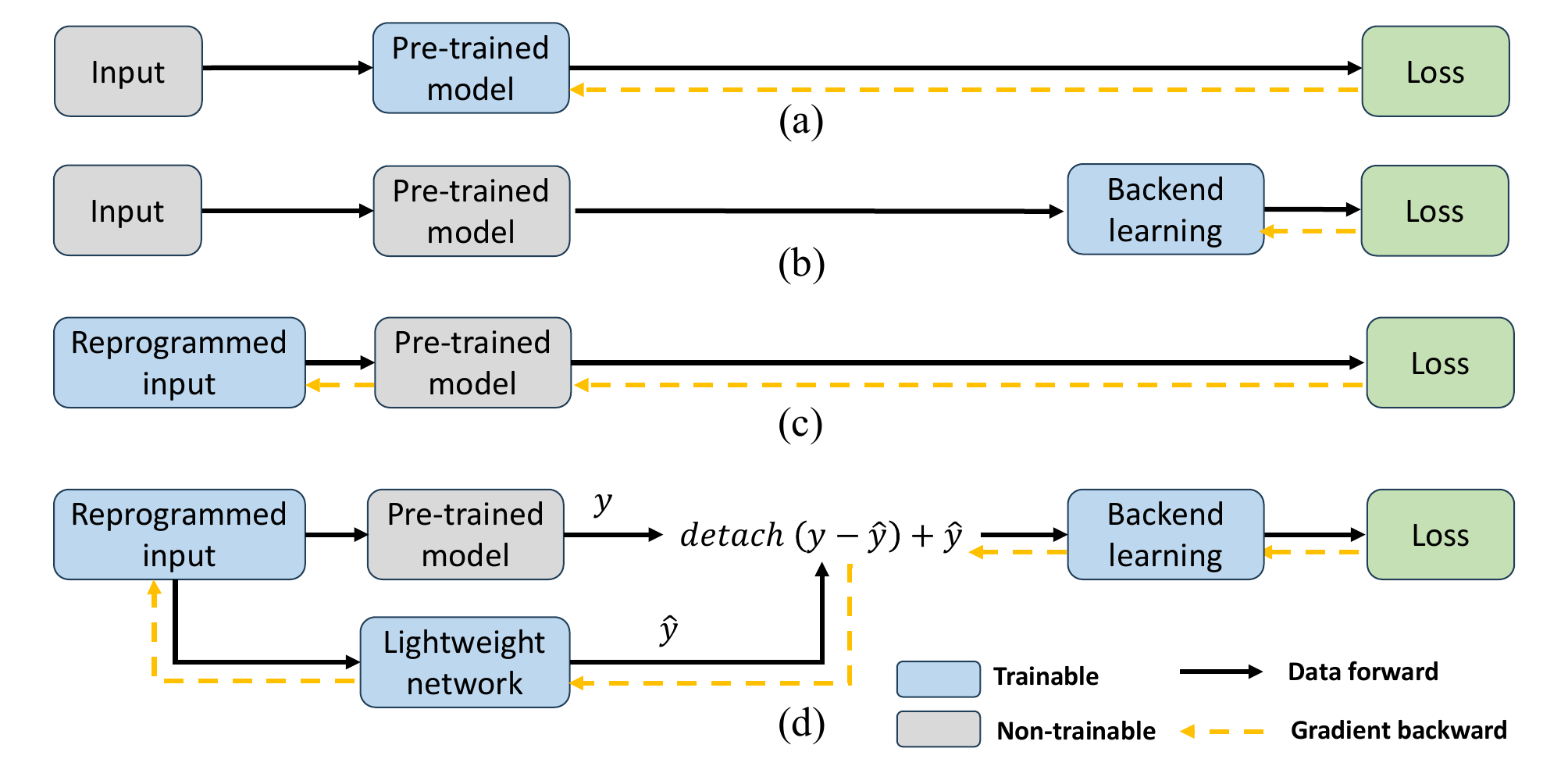}
  }
  \caption{(a) Model finetune, (b) Backend learning, (c) Reprogramming, (d) The proposed model}
  \label{fig:network}
  \vspace{-2mm}
\end{figure}

The obvious advantage of reprogramming is that it can transfer a pre-trained model for different tasks by only modifying the input without finetuning the model. This is crucial for resource efficiency, as the computation cost is high in training or finetuning large-size DNN models. However, \cite{sung2022lst} argued that it may not be efficient during training. To update the parameters at the input level, the gradient backpropagation has to go through the whole pre-trained model to reach the top of the model, as shown in Fig.~\ref{fig:network}(c). It costs a large amount of memory to save the storage of intermediate activation.

Besides, vanilla reprogramming is not suitable for black-box models. To address this problem, \cite{tsai2020transfer} proposed to utilize zeroth order optimization to estimate the gradient for the learnable parameters at the model input. This work shows that reprogramming can be effective in transfer learning for black-box models for medical data. In each update step of the model input in this work, the estimation of the gradient requires more than 10 times forward steps, which is not efficient. 




\section{Methodology}
\label{sec:methodology}
 
This paper focuses on: $(1)$ adapting black-box SV models with reprogramming and backend learning for language mismatch scenarios and $(2)$ high-efficiency learning. Without specifying, the pre-trained black-box model is the speaker embedding extraction model in SV. The proposed work is close to \cite{tsai2020transfer}, but a different method for gradient estimation is utilized here. The whole system pipeline is shown as Fig.~\ref{fig:network}(d).

\subsection{Input Reprogramming}
\label{ssec:in_reprog}
The speech input data is reprogrammed on the waveform level as \cite{yang2021voice2series}. A sequence of learnable parameters $\mathbf{W}=[w_1, w_2, ..., w_n]$ is concatenated on both sides of the original waveform $\mathbf{X}=[x_1, x_2, ..., x_l]$ as:
\begin{equation}
    \tilde{\mathbf{W}}=[w_1, ...w_{n//2}, x_1, ..., x_l, w_{n//2+1}, ..., w_n]
  \label{eq:concat}
\end{equation}
The reprogrammed waveform is transformed into a spectrum as the input for the speaker embedding extraction model.

\subsection{Gradient Estimation Network}
\label{ssec:network}
In this work, we propose to add a siamese neural network parallel with the original pre-trained model for the purpose of gradient estimation, as shown in Fig.~\ref{fig:network}(d). The pre-trained model is denoted as $F$ and the siamese network is represented 
as $\hat{F}_{\theta}$, where $\theta$ is the learnable weight of this network. 
The adaptation output is given as:
\begin{equation*}
    y = F(\Tilde{x})\ \ \ ,\ \ \ \ \hat{y} = \hat{F}_{\theta}(\Tilde{x})\ \ \ 
  \label{eq:model1}
\end{equation*}
\begin{equation}
    y_{adapt} = detach(y - \hat{y}) + \hat{y}\ \ \ 
  \label{eq:adapt}
\end{equation}
 where $\tilde{x}$ is the reprogrammed model input described in Section \ref{ssec:in_reprog} and $detach()$ is an operation to stop the gradient. $y$ represents the extracted speaker embeddings from the pre-trained model. The value of $y_{adapt}$ equals the original output $y$, which gives two advantages: 

 $(1)$ The gradient can bypass the black-box model to update the learnable input $\mathbf{W}$. The optimization targets for $y$ and $y_{adapt}$ are the same. In other words, defining the objective function as $L(y, spk_i)$, e.g., cross-entropy classification loss, we have:
\begin{equation}
   \argmin_{\Tilde{x}} L(y, spk_i) = \argmin_{\Tilde{x}} L(y_{adapt}, spk_i)
  \label{eq:argmin}
\end{equation}
where $spk_i$ represents the target speaker with index $i$.
The detach operation stops the gradient calculation in $(y - \hat{y})$, and the gradient backpropagation on $y$ passes through $\hat{F}_{\theta}$ to update $\tilde{x}$. By optimizing the objective function $L(y_{adapt}, spk_i)$, the loss on $y$ is also decreased. In a special case, if $\hat{F}_{\theta}$ is exactly the same as $F$, i.e., $F(x)=\hat{F}_{\theta}(x)$ for all $\mathbf{X}$, the gradient estimated by $\hat{F}_{\theta}$ is nothing different from the gradient calculated through $F$. By designing a lightweight $\hat{F}_{\theta}$, the computation cost for gradient backpropagation is decreased.

$(2)$ $\hat{F}_{\theta}$ is merely utilized for gradient estimation, and it is abandoned after training, which reduces the size of the adaption model. The structure of the proposed adaption system is similar to the structure of Rep-Net\cite{yang2022rep}. \cite{yang2022rep} also builds a lightweight branch network parallel with the pre-trained model, but it is utilized for intermediate feature exchange with the pre-trained model and feature fusion at the output stage. Thus Rep-Net requires more parameters for the final model.

  \subsection{Backend Learning}
\label{ssec:backend}
The backend learning module takes the output $y_{adapt}$ from the previous module as input and produces the adapted speaker embedding. This module is designed to be simple for the purpose of adding fewer parameters and less model complexity, which is also in favor of avoiding model overfitting. 

Two types of networks are utilized in this work. Previous work \cite{li22m_interspeech} has shown that the speaker embeddings' statistics information, e.g., mean and standard deviation, contains domain bias. By normalizing the embeddings with source-domain statistics, the domain mismatch can be alleviated. The first type of backend learning module is only a batch normalization (BN) layer, denoted as $Back_{BN}$.

The second type of backend learning module consists of two fully connected (FC) layers and a residual connection. There is a BN layer and a ReLU activation function between the two FC layers. This type of module is indicated as $Back_{FC-k}$, where $k$ denotes the number of neural nodes in the middle, controlling the complexity of this module. 

 \subsection{Objective Function}
\label{ssec:loss}
In vanilla reprogramming for classification tasks, the model outputs are mapped from the source domain on target-domain labels\cite{elsayed2018adversarial}, e.g., mapping a class label in ImageNet on a digit in MNIST. In SV, the speaker labels for the pre-trained model are not available. Thus we simply trained the outputs with a speaker classifier layer. The AAM-softmax\cite{deng2019arcface} with a margin of $0.3$ and scale of $20$ is utilized as the objective function on the output speaker embeddings. 

\begin{table*}[t]
  \caption{Performances of models on the test set of CN-Celeb. The model size is given below the model name. We represent the number of parameters involved in gradient backpropagation (\textbf{Para. BP})  by the percentage over the original model size, \textbf{Para. Add} stands for the number of parameters added after adaptation training. The parameters in the classification layer are ignored. Reprog. is short for vanilla reprogramming as Fig.~\ref{fig:network}(c). Grad.Reprog. stands for the proposed gradient estimated reprogramming. $C=16$, $n=4800$.}
  \vspace{1mm}
  \label{tab:result}
  \centering
  \scalebox{1}{
  \begin{tabular}{lllllc}
    \toprule
     \textbf{Model (Size)}             & \textbf{Adaptation}       & \textbf{EER(\%)}  & \textbf{Para. BP($\%$)}   & \textbf{Para. Add($\%$)}  & \textbf{Black-box training allow}\\
    \midrule
        \midrule
    \multirow{14}{*}{\shortstack[l]{SE-\\ResNet\\34\\(6.98M)}}     & None        & 11.5    &  0       & 0    & \checkmark \\
                                  & Full-finetune                       & 8.83    & 100   & 0    & \ding{53} \\
                                 & $Back_{BN}$                          & 9.3     & 0.007      & 0.007  & \checkmark     \\
                                 & $Back_{FC-32}$                        & 8.63    & 0.262     & 0.262  & \checkmark  \\     
                                 & $Back_{FC-64}$                        & 8.56    & 0.498     & 0.498  & \checkmark\\
                                 & $Back_{FC-128}$                       & 8.37    & 0.97     &  0.97  & \checkmark \\   
                                 & $Back_{FC-256}$                       & 8.3     & 1.91    & 1.91  & \checkmark\\
                                 & Reprog.+$Back_{BN}$             & 10.16   & 100.076   & 0.076 & \ding{53} \\
                                 & Reprog.+$Back_{FC-16}$          & 8.2        & 100.212   & 0.212 & \ding{53} \\   
                                 & Reprog.+$Back_{FC-32}$          & 8.01    & 100.331   & 0.331 & \ding{53} \\
                                 & Reprog.+$Back_{FC-64}$          & 8.26    & 100.567   & 0.567 & \ding{53} \\    
                                 & Grad.Reprog.+$Back_{FC-16}$    & 8.32    & 0.924     & 0.212 & \checkmark     \\   
                                 & Grad.Reprog.+$Back_{FC-32}$    & 8.61    & 1.041     & 0.331 & \checkmark     \\
                                 & Grad.Reprog.+$Back_{FC-64}$    & \textbf{7.91}    & 1.275       & 0.567 & \checkmark     \\                   
    \midrule
    \multirow{14}{*}{\shortstack[l]{ECAPA-\\TDNN\\512\\(5.95M)}}     & None        & 17.78   &  0       & 0   & \checkmark \\
                                  & Full-finetune                     & 9.56    & 100   & 0   & \ding{53} \\  
                                 & $Back_{BN}$                    & 11.02   & 0.009      & 0.009 & \checkmark \\
                                 & $Back_{FC-32}$                  & 10.13   & 0.307     & 0.307 & \checkmark \\    
                                 & $Back_{FC-64}$                  & 10.39   & 0.584     & 0.584 & \checkmark \\
                                 & $Back_{FC-128}$                 & 10.2    & 1.137     & 1.137 & \checkmark \\  
                                 & $Back_{FC-256}$                 & 9.71    & 2.244    &  2.244 & \checkmark \\
                                 & Reprog.+$Back_{BN}$             & 10.43   & 100.089   & 0.089 & \ding{53} \\  
                                 & Reprog.+$Back_{FC-16}$          & 9.53    & 100.249   & 0.249 & \ding{53} \\      
                                 & Reprog.+$Back_{FC-32}$          &  \textbf{9.29}    & 100.388     & 0.388 & \ding{53} \\  
                                 & Reprog.+$Back_{FC-64}$          & 9.44    & 100.665   & 0.665 & \ding{53} \\  
                                 & Grad.Reprog.+$Back_{FC-16}$    & 9.78    & 1.083     & 0.249 & \checkmark \\
                                 & Grad.Reprog.+$Back_{FC-32}$    & 9.99    & 1.221     & 0.388 & \checkmark \\
                                 & Grad.Reprog.+$Back_{FC-64}$    & 9.94    & 1.495       & 0.665 & \checkmark \\
    \bottomrule
  \end{tabular}}
  \vspace{-3mm}
\end{table*}

\section{Experimental settings}
\label{sec:experiment}

\subsection{Datasets}
\label{ssec:data}
The datasets utilized in our experiments are VoxCeleb 1 \& 2\cite{nagrani2017voxceleb,chung2018voxceleb2,nagrani2020voxceleb} and CN-Celeb 1\cite{fan2020cn}. All data are $16k$Hz records. VoxCeleb 2 is used for model pre-training, which is a large-scale SV dataset, consisting of $5,994$ speakers. There are more than $1$ million utterances in this dataset and most of the utterances are spoken in English.

CN-Celeb is a challenging Chinese SV dataset. The pre-trained models have a large domain mismatch on this dataset. The pre-trained model is adapted using the training set of CN-Celeb and evaluated on its evaluation set. There are around $800$ and $200$ speakers in the training set and evaluation set, respectively.

\subsection{Network structure}
\label{ssec:backbone}
Two pre-trained models are used in this work, ECAPA-TDNN-512\cite{desplanques20_interspeech} and SE-ResNet34\cite{hu2018squeeze,qi2020deep} with 32 planes. The dimension of extracted speaker embedding is $256$. They are two commonly used networks for speaker embedding extraction. The number of parameters in ECAPA and SE-ResNet is 5.95 and 6.98 million ($M$), respectively. Their Equal Error Rate (EER) on Vox.O is $1.12\%$ and $1.3\%$.

The gradient estimation network used here is ECAPA-TDNN with channel $C$. Different values of $C$ are evaluated in the experiments.

\subsection{Model Training}
\label{ssec:traiing}
Here we only describe the model training in the adaptation process. The batch size is set to $128$. In each training step, a 2-second duration segment is randomly cropped from each input utterance. The raw waveform of the cropped segment is reprogrammed as Section~\ref{ssec:in_reprog} and transformed into 64-dimension log Mel-filterbanks (FBank) as model input. The learnable input parameters $\mathbf{W}$ are initialized as $0$, which simulates silence addition with the original waveform $\mathbf{X}$. The adaption model is trained by the Adam optimizer with a weight decay of $1e-4$. The learning rate is initialized as $1e-3$, and reduced by a ratio of $10$ at the $10_{th}$ and $15_{th}$ epoch, respectively. The training is stopped after $20$ epochs.

\section{Results and Discussion}
\label{sec:results}
The results are given in Table~\ref{tab:result}. The length of learnable parameters $\mathbf{W}$ is set to equal $0.3s$ duration of a waveform, i.e., $n=4800$ for $16k$ Hz input audio. Considering the efficiency of the model process, we do not choose a very large $n$. Another reason is the task gap in adaptation is smaller than the tasks in other transfer learning work(voice2series). Adaptation does not require generating significant modification on outputs. The channel $C$ of the gradient estimation network is set to be $16$ in Table~\ref{tab:result}.

\subsection{Baselines}
\label{ssec:baseline}
The full-model finetune and backend learning are the baselines. Full finetune observably decreases the EER in the adaptation models, while it consumes numerous computation costs during training. $Back_{FC}$ shows close or better results than the full-finetuned model in SE-ResNet34, with significantly fewer parameters involved during gradient backpropagation. With a larger value of $C$, the performance of $Back_{FC}$ is improved. On ECAPA-TDNN 512, the fully finetuned model outperforms the backend learning models.

\subsection{Vanilla Reprogramming}
\label{ssec:vanilla_reprog}
We combine the vanilla reprogramming (Fig.~\ref{fig:network}(c)) with different backend modules and evaluate their effects. The gradient backpropagation goes through the pre-trained model to update the $\mathbf{W}$ at the input. The reprogramming with $Back_{BN}$ gives no superior results than other backend learning adaptations, which indicates that using merely a small number of learnable $\mathbf{W}$ cannot modify the output significantly enough for adaption. By involving reprogramming with $FC$, the adapted models' performance is boosted notably and outperforms the full finetune adaptation approach. After training, the number of addition parameters is small. However, the vanilla reprogramming leads to a large number of parameters in the gradient backpropagation during training.

\subsection{Gradient Estimated Reprogramming}
\label{ssec:grad_reprog}
The proposed reprogramming with a backend learning module outperforms the fully finetuned model on SE-ResNet34 with much less computation cost for training. When using the same number of parameters, the proposed method shows better results than the raw backend learning module. The proposed reprogramming is surpassed by the vanilla reprogramming on both models generally, which is reasonable because it is updated using the gradient estimation network. The number of additional parameters and parameters involved in training is small, which gives both parameter and memory efficiency.

We evaluate the effect of the proposed method using different learnable input lengths ($n$ in Section~\ref{ssec:in_reprog}) and give the results in Table.~\ref{tab:lenght}. With a longer learnable input $\mathbf{W}$, the EER is decreased in SE-ResNet34, which indicates that the $\mathbf{W}$ with a larger size produces more information to improve the adaptation. While this factor does not affect the results on ECAPA-TDNN a lot. How to choose a good $n$ for better adaptation results requires more study in the future.

Table~\ref{tab:C_length} shows the results of the proposed model using different $C$ for the gradient estimation network. The EER is increased slightly when increasing the value of $C$ on ECAPA-TDNN512. $C=16$ gives the best result for SE-ResNet34. This may indicate that estimating the gradient for updating $\mathbf{W}$ does not have to be very complex, which can decrease the computation cost during training.

\begin{table}[t]
  \caption{Performances on CN-Celeb with different lengths of $\mathbf{W}$. The unit of the length used here is second (s). $C=16$ for gradient estimation network. $Back_{FC-64}$ is utilized.}
  \vspace{1mm}
  \label{tab:lenght}
  \centering
 \setlength\tabcolsep{7pt}
  \scalebox{0.96}{
  \begin{tabular}{lccc}
    \toprule
    \textbf{Model}   & \textbf{0.2s}    & \textbf{0.3s}   & \textbf{0.5s}\\  
    \midrule
    \midrule
    SE-ResNet34      & 8.1             & 7.91           &  7.87      \\
    \midrule
    ECAPA-TDNN512   & 9.95            & 9.94           &  10.06     \\
    \bottomrule
  \end{tabular}}
  \vspace{-3.5mm}
\end{table}

\begin{table}[t]
  \caption{Performances on CN-Celeb with different $C$. The length of $\mathbf{W}$ is $0.3$ second. $Back_{FC-64}$ is utilized.}
  \vspace{1mm}
  \label{tab:C_length}
  \centering
 \setlength\tabcolsep{7pt}
  \scalebox{0.96}{
  \begin{tabular}{lccccc}
    \toprule
    \textbf{Model}   & \textbf{4}    & \textbf{8}   & \textbf{16}    & \textbf{32}   & \textbf{64} \\  
    \midrule
    \midrule
    SE-ResNet34     &  8.41        &  8.18        &   7.91         & 8.26          & 8.35    \\
    \midrule
    ECAPA-TDNN512   & 9.79          & 9.8          &  9.94          & 9.94          & 10.31     \\
    \bottomrule
  \end{tabular}}
  \vspace{-3.5mm}
\end{table}

\section{Conclusion}
\label{sec:print}
In this work, we propose an efficient domain adaption method for black-box SV models. The adaption method involves a modified version of the reprogramming algorithm augmented with backend learning and a lightweight network. The lightweight network is built parallel to the pre-trained model and utilized for gradient backpropagation to the learnable input bypassing the black-box model. The proposed method achieves performance close to or surpassing the fully finetuned adaptation models. A small number of parameters is utilized during training and few new parameters are added to the adaptation model, which achieves both efficiency in parameters and computation memory. 





\vfill\pagebreak

\bibliographystyle{IEEEbib}
\bibliography{strings,refs}

\begin{thebibliography}{10}

\bibitem{liu2015deep}
Yuan Liu, Yanmin Qian, Nanxin Chen, Tianfan Fu, Ya~Zhang, and Kai Yu,
\newblock ``Deep feature for text-dependent speaker verification,''
\newblock {\em Speech Communication}, vol. 73, pp. 1--13, 2015.

\bibitem{xie2019utterance}
Weidi Xie, Arsha Nagrani, Joon~Son Chung, and Andrew Zisserman,
\newblock ``Utterance-level aggregation for speaker recognition in the wild,''
\newblock in {\em ICASSP}. IEEE, 2019, pp. 5791--5795.

\bibitem{chung20b_interspeech}
Joon~Son Chung, Jaesung Huh, Seongkyu Mun, Minjae Lee, Hee-Soo Heo, Soyeon
  Choe, Chiheon Ham, Sunghwan Jung, Bong-Jin Lee, and Icksang Han,
\newblock ``{In Defence of Metric Learning for Speaker Recognition},''
\newblock in {\em Interspeech}, 2020, pp. 2977--2981.

\bibitem{10094954}
Bing Han, Zhengyang Chen, and Yanmin Qian,
\newblock ``Exploring binary classification loss for speaker verification,''
\newblock in {\em ICASSP}, 2023, pp. 1--5.

\bibitem{snyder2017deep}
David Snyder, Daniel Garcia-Romero, Daniel Povey, and Sanjeev Khudanpur,
\newblock ``Deep neural network embeddings for text-independent speaker
  verification.,''
\newblock in {\em Interspeech}, 2017, pp. 999--1003.

\bibitem{snyder2018x}
David Snyder, Daniel Garcia-Romero, Gregory Sell, Daniel Povey, and Sanjeev
  Khudanpur,
\newblock ``X-vectors: Robust dnn embeddings for speaker recognition,''
\newblock in {\em ICASSP}. IEEE, 2018, pp. 5329--5333.

\bibitem{desplanques20_interspeech}
Brecht Desplanques, Jenthe Thienpondt, and Kris Demuynck,
\newblock ``{ECAPA-TDNN: Emphasized Channel Attention, Propagation and
  Aggregation in TDNN Based Speaker Verification},''
\newblock in {\em Interspeech}, 2020, pp. 3830--3834.

\bibitem{lin2020framework}
Weiwei Lin, Man-Wai Mak, Na~Li, Dan Su, and Dong Yu,
\newblock ``A framework for adapting dnn speaker embedding across languages,''
\newblock {\em IEEE/ACM Transactions on Audio, Speech, and Language
  Processing}, vol. 28, pp. 2810--2822, 2020.

\bibitem{wang2018unsupervised}
Qing Wang, Wei Rao, Sining Sun, Leib Xie, Eng~Siong Chng, and Haizhou Li,
\newblock ``Unsupervised domain adaptation via domain adversarial training for
  speaker recognition,''
\newblock in {\em ICASSP}. IEEE, 2018, pp. 4889--4893.

\bibitem{chen2020adversarial}
Zhengyang Chen, Shuai Wang, and Yanmin Qian,
\newblock ``Adversarial domain adaptation for speaker verification using
  partially shared network.,''
\newblock in {\em Interspeech}, 2020, pp. 3017--3021.

\bibitem{alam2018speaker}
Md~Jahangir Alam, Gautam Bhattacharya, and Patrick Kenny,
\newblock ``Speaker verification in mismatched conditions with frustratingly
  easy domain adaptation.,''
\newblock in {\em Odyssey}, 2018, vol. 2018, pp. 176--180.

\bibitem{li22m_interspeech}
Jingyu Li, Wei Liu, and Tan Lee,
\newblock ``{EDITnet: A Lightweight Network for Unsupervised Domain Adaptation
  in Speaker Verification},''
\newblock in {\em Interspeech}, 2022, pp. 3694--3698.

\bibitem{rohdin2019speaker}
Johan Rohdin, Themos Stafylakis, Anna Silnova, Hossein Zeinali, Luk{\'a}{\v{s}}
  Burget, and Old{\v{r}}ich Plchot,
\newblock ``Speaker verification using end-to-end adversarial language
  adaptation,''
\newblock in {\em ICASSP}. IEEE, 2019, pp. 6006--6010.

\bibitem{ren2015faster}
Shaoqing Ren, Kaiming He, Ross Girshick, and Jian Sun,
\newblock ``Faster r-cnn: Towards real-time object detection with region
  proposal networks,''
\newblock {\em Advances in neural information processing systems}, vol. 28,
  2015.

\bibitem{hsu2021hubert}
Wei-Ning Hsu, Benjamin Bolte, Yao-Hung~Hubert Tsai, Kushal Lakhotia, Ruslan
  Salakhutdinov, and Abdelrahman Mohamed,
\newblock ``Hubert: Self-supervised speech representation learning by masked
  prediction of hidden units,''
\newblock {\em IEEE/ACM Transactions on Audio, Speech, and Language
  Processing}, vol. 29, pp. 3451--3460, 2021.

\bibitem{li2022coral++}
Rongjin Li, Weibin Zhang, and Dongpeng Chen,
\newblock ``The coral++ algorithm for unsupervised domain adaptation of speaker
  recognition,''
\newblock in {\em ICASSP}. IEEE, 2022, pp. 7172--7176.

\bibitem{elsayed2018adversarial}
Gamaleldin~F Elsayed, Ian Goodfellow, and Jascha Sohl-Dickstein,
\newblock ``Adversarial reprogramming of neural networks,''
\newblock in {\em International Conference on Learning Representations}, 2018.

\bibitem{tsai2020transfer}
Yun-Yun Tsai, Pin-Yu Chen, and Tsung-Yi Ho,
\newblock ``Transfer learning without knowing: Reprogramming black-box machine
  learning models with scarce data and limited resources,''
\newblock in {\em ICML}. PMLR, 2020, pp. 9614--9624.

\bibitem{yang2021voice2series}
Chao-Han~Huck Yang, Yun-Yun Tsai, and Pin-Yu Chen,
\newblock ``Voice2series: Reprogramming acoustic models for time series
  classification,''
\newblock in {\em ICML}. PMLR, 2021, pp. 11808--11819.

\bibitem{hung2023low}
Yun-Ning Hung, Chao-Han~Huck Yang, Pin-Yu Chen, and Alexander Lerch,
\newblock ``Low-resource music genre classification with cross-modal neural
  model reprogramming,''
\newblock in {\em ICASSP}. IEEE, 2023, pp. 1--5.

\bibitem{yang2023english}
Chao-Han~Huck Yang, Bo~Li, Yu~Zhang, Nanxin Chen, Rohit Prabhavalkar, Tara~N
  Sainath, and Trevor Strohman,
\newblock ``From english to more languages: Parameter-efficient model
  reprogramming for cross-lingual speech recognition,''
\newblock in {\em ICASSP}. IEEE, 2023, pp. 1--5.

\bibitem{yen23_interspeech}
Hao Yen, Pin-Jui Ku, Chao-Han~Huck Yang, Hu~Hu, Sabato~Marco Siniscalchi,
  Pin-Yu Chen, and Yu~Tsao,
\newblock ``{Neural Model Reprogramming with Similarity Based Mapping for
  Low-Resource Spoken Command Recognition},''
\newblock in {\em INTERSPEECH}, 2023, pp. 3317--3321.

\bibitem{sung2022lst}
Yi-Lin Sung, Jaemin Cho, and Mohit Bansal,
\newblock ``Lst: Ladder side-tuning for parameter and memory efficient transfer
  learning,''
\newblock {\em Advances in Neural Information Processing Systems}, vol. 35, pp.
  12991--13005, 2022.

\bibitem{yang2022rep}
Li~Yang, Adnan~Siraj Rakin, and Deliang Fan,
\newblock ``Rep-net: Efficient on-device learning via feature reprogramming,''
\newblock in {\em CVPR}, 2022, pp. 12277--12286.

\bibitem{deng2019arcface}
Jiankang Deng, Jia Guo, Niannan Xue, and Stefanos Zafeiriou,
\newblock ``Arcface: Additive angular margin loss for deep face recognition,''
\newblock in {\em CVPR}, 2019, pp. 4690--4699.

\bibitem{nagrani2017voxceleb}
Arsha Nagrani, Joon~Son Chung, and Andrew Zisserman,
\newblock ``Voxceleb: A large-scale speaker identification dataset,''
\newblock {\em Interspeech}, pp. 2616--2620, 2017.

\bibitem{chung2018voxceleb2}
Joon~Son Chung, Arsha Nagrani, and Andrew Zisserman,
\newblock ``Voxceleb2: Deep speaker recognition,''
\newblock {\em Interspeech}, pp. 1086--1090, 2018.

\bibitem{nagrani2020voxceleb}
Arsha Nagrani, Joon~Son Chung, Weidi Xie, and Andrew Zisserman,
\newblock ``Voxceleb: Large-scale speaker verification in the wild,''
\newblock {\em Computer Speech \& Language}, vol. 60, pp. 101027, 2020.

\bibitem{fan2020cn}
Yue Fan, JW~Kang, LT~Li, KC~Li, HL~Chen, ST~Cheng, PY~Zhang, ZY~Zhou, YQ~Cai,
  and Dong Wang,
\newblock ``Cn-celeb: a challenging chinese speaker recognition dataset,''
\newblock in {\em ICASSP}. IEEE, 2020, pp. 7604--7608.

\bibitem{hu2018squeeze}
Jie Hu, Li~Shen, and Gang Sun,
\newblock ``Squeeze-and-excitation networks,''
\newblock in {\em CVPR}, 2018, pp. 7132--7141.

\bibitem{qi2020deep}
Minhui Qi, Yongbin Yu, Yifan Tang, QuanXin Deng, Feng Mai, and Nima Zhaxi,
\newblock ``Deep cnn with se block for speaker recognition,''
\newblock in {\em ICTC}. IEEE, 2020, pp. 240--244.

\end{thebibliography}

\end{document}